# Switching on Antiferroelectrics

G. Catalan,[1,2] A. Gruverman,[3] J. Íñiguez-González,[4,5] D. Meier,[6] and M. Trassin[7]

Correspondence to gustau.catalan@icn2.cat, agruverman2@unl.edu, jorge.iniguez@list.lu, dennis.meier@ntnu.no, and morgan.trassin@mat.ethz.ch

[1] ICREA-Institucio Catalana de Recerca I Estudis Avançats, Barcelona, Catalonia
[2] ICN2-Institut Catala de Nanociencia I Nanotecnologia, BIST-CSIC-UAB, Bellaterra, Catalonia
[3] Department of Physics and Astronomy, University of Nebraska, Lincoln, NE 68588-0299, USA
[4] Smart Materials Unit, Luxembourg Institute of Science and Technology (LIST), L-4362 Esch-sur-Alzette, Luxembourg
[5] Department of Physics and Materials Science, University of Luxembourg, L-4422 Belvaux, Luxembourg
[6] Department of Materials Science and Engineering, NTNU Norwegian University of Science and Technology, Trondheim, Norway
[7] Department of Materials, ETH Zurich, Vladimir-Prelog-Weg 4, Zurich 8093, Switzerland

**Antiferroelectrics attract broad attention due to their unusual physical characteristics, chief among which is the double-hysteresis loop that separates their antipolar ground state from the voltage-induced polar phase, which is promising for applications in energy storage and electrocaloric cooling. However, their defining features (antipolar ground state and double-hysteresis loops) are increasingly challenged: materials with non-collinear and/or hybrid polar-antipolar order have been discovered, and double-hysteresis has been realized in materials without a conventional antipolar ground state. These developments add to the intensifying interest in fundamental and practical aspects of antiferroelectrics, and call for a fresh look at antiferroelectricity. In this Perspective, we provide an updated and all-encompassing definition of antiferroelectricity, discuss material systems with new antipolar orders and/or engineered double hysteresis, and reflect on emergent properties and theoretical approaches. This work casts a bird's eye view on the rapidly evolving trends that are shaping up the research on ferroics with antipolar order.**

## FUNDAMENTALS

Although in their antipolar ground state antiferroelectrics lack net polarization, this does not make them scientifically or technologically less appealing than their ferroelectric counterparts. The emerging interest in antiferroelectrics is driven by their unique physical properties (in particular, their field-induced transition to a polar phase) that make them suitable for high-density energy storage, resistive memory, electromechanical actuation and electrocaloric applications[1–8]. These applications have been covered by thorough reviews published over the last several years and we will not dwell on them[9–11]. Instead, we present a perspective on antiferroelectricity that accounts for the rapid and multifaceted developments in the field, focusing on newly discovered material systems with antipolar order, along with emergent phenomena, theoretical approaches, and evolving trends in the field of antiferroelectricity.

### History and Classical Definition

Antiferroelectrics started as a theoretical idea of Kittel[12], based on the intuitive notion that, if antiparallel order can exist for magnetic spins (what we call antiferromagnetism), it should also exist for electric dipoles. Applying an external electric field to the dipoles could then result in their realignment from an antiparallel state to a polarized phase, with a double hysteresis loop between the antipolar phase and the two polarities of the field-induced polar one. He wrote his phenomenological theory of antiferroelectricity in 1951 and, within the same year, the first experimental evidence for antiferroelectricity was provided by Shirane and co-workers, who reported double-hysteresis loops in $PbZrO_3$[13].

Despite the intuitive nature of the concept, a rigorous definition of antiferroelectricity has turned out to be difficult, for various reasons. One is that, even if we accept that all antiferroelectrics should display double hysteresis loops, not all materials with double hysteresis are antiferroelectric. For example, ferroelectrics with a first-order phase transition, such as $BaTiO_3$, can display double hysteresis when the paraelectric-ferroelectric transition is induced by a field just above $T_C$. Such transition is not between an antipolar and a polar state, but between a strictly non-polar (cubic, in the case of $BaTiO_3$) and a polar phase. The root of the problem here is that the free energy for both antiferroelectrics and first-order ferroelectrics is a triple well, with the central minimum being antipolar for antiferroelectrics and non-polar for "normal" ferroelectrics. In addition, defect dipoles or aging can cause a "pinching" of ferroelectric hysteresis loops that looks antiferroelectric-like[14]. The problem of correlating double hysteresis with antiferroelectricity has come to the fore, as a double-hysteretic transition to a polar phase has been found in $ZrO_2$ thin films[15], the end member of the ferroelectric hafnia-zirconia family, for which the antipolar requisite of the ground state is under scrutiny[16,17].

The second difficulty is deciding what counts as *antipolar*, because *any* ionic crystal can be viewed as an antiparallel arrangement of dipoles. By way of illustration, consider a one-dimensional ionic chain: $+-+-\ldots$, which can be seen as an antipolar arrangement where dipoles alternate direction $+\leftarrow-\rightarrow+\leftarrow-\ldots$, yet few would call this structure antipolar, despite the fact that charged atomic planes can cause an interfacial "polar catastrophe"[18,19]. It is worth to recall that defining absolute polarization has historically been troublesome in ferroelectrics, where a rigorous treatment involves the transient macroscopic current associated with the onset of the polar distortion[20,21]; this notion has no obvious analog for antipolar displacements, which yield no net electric current. Such difficulties do not afflict magnetic materials because spins are local (and quantized) entities that do not require relative changes (or switching) to be defined. As Tolédano and Guennou[22] stated, "a paraelectric *state* does not exist by itself in the same sense as the paramagnetic state, and all crystal structures displaying nonpolar symmetries can be potentially antiferroelectric".

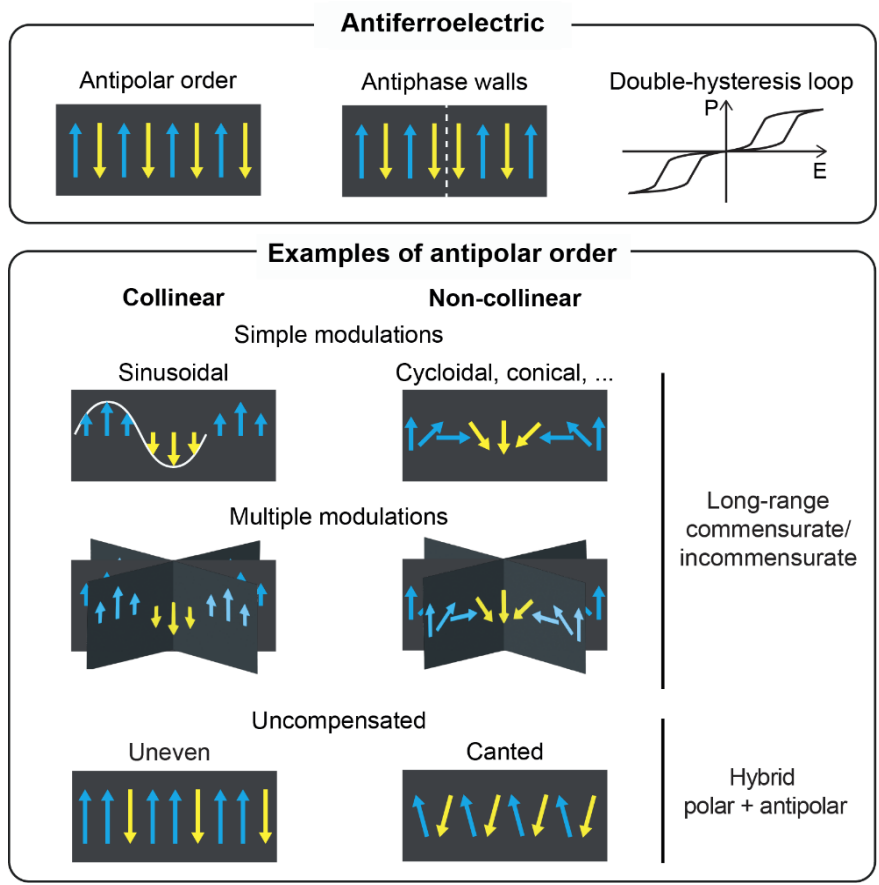

Fig. 1 | **Antiferroelectricity and antipolar order of electric dipoles**. (Upper panel) Common characteristics of antiferroelectrics: modulated antipolar electric dipoles (illustrated by blue and yellow arrows), antiphase (translational) boundaries / domain walls, and double-hysteresis. (Lower panel) In general, electric dipoles can form complex collinear and non-collinear textures, including both macroscopically compensated (P = 0) and uncompensated (P ≠ 0) arrangements. Examples for compensated collinear textures encompass classical antipolar order, as well as Ising-like modulated order (sinusoidal). Furthermore, simple one-dimensional (1D) or multiple higher-dimensional (2D or 3D) textures with P = 0 arise, ranging from Bloch- and Néel-like order to dipolar skyrmions or hopfions. Partially compensated dipolar order with unequal polarization contributions exists in ferrielectric (uneven) and canted antipolar systems.

The difference between dipoles and spins has even led eminent figures to express their skepticism about the concept of antiferroelectricity. Landau & Lifshitz stated in their textbook[23] that "There are no crystals without *electric structure*, and therein lies the essential difference between the electric and the magnetic properties of crystals". Arkady Levanyuk was more direct: "The concept of an antiferroelectric state turns out to be superfluous in general"[24]. Jim Scott, in his famous blunt style, wrote that "antiferroelectricity is an ill-defined, almost useless concept"[25].

We take a more sanguine view and, alongside some others[22,26], believe that antiferroelectricity is a useful and physically meaningful concept that complements ferroelectricity. As in ferroelectrics, one can define individual dipoles as atomic displacements with reference to a higher symmetry phase. The simplest may be a displacement of half of the ions in one direction and the other half in the opposite, thereby doubling the size of the antiferroelectric unit cell with respect to the paraelectric one. Despite his reservations about antiferroelectrics, Scott did in fact provide a definition along those lines[25]: "An antiferroelectric is a crystal having a polar optical phonon whose eigenvector is the same as the ionic displacements required to transform the lattice to a centric structure having half as many ions per primitive cell". Moreover, as in ferroelectrics, antiferroelectrics can have mobile boundaries between domains, and this is a distinctive feature of antipolar materials vs non-polar ones.

Nevertheless, it is now clear that more complex (anti)polar modulations are possible[27], as illustrated in Fig. 1. In addition, antiferroelectric-like functionalities, in particular double hysteresis, are being engineered using non-antipolar materials. It therefore seems useful to re-examine the definition of antiferroelectrics.

**A Modern Definition**

Based on the above considerations, a general definition of antiferroelectrics could be provided as follows:

"Antiferroelectrics are crystalline materials, which (i) have a structure characterized by a periodic antipolar arrangement of electric dipoles arising due to atomic displacements with reference to a higher symmetry phase, (ii) are able to accommodate antiphase (translational) boundaries in the periodic modulation of said polar displacements, and (iii) can undergo a reversible electric-field-induced transition from the antipolar to a polar phase, manifested as a double hysteresis polarization loop."

This definition assumes a spin-like picture where the field-driven transition to the polar state involves a reorientation of polar distortions already present in the antiferroelectric state. With this, we want to distinguish between systems widely regarded as antiferroelectrics, such as $PbZrO_3$, where the dipoles arise due to Pb-O displacements both for the antiferroelectric and the ferroelectric phase, from the more common situation in perovskites with octahedral tilts, where the alternating sign of the tilts may be viewed as antipolar displacements, but whose eventual field-induced transition to a polar state involves a different set of dipoles. Another way to put it is that the antipolar and polar distortions should have the same atomistic nature or, more technically, belong to the same phonon band.

There are in fact many systems that meet the proposed definition only partially, as illustrated in the Venn diagram in Fig. 2. Here, we will refer to these materials as **pseudo antiferroelectrics** (although in the literature they are also indistinctly called *antiferroelectric-like, quasi-antiferroelectric*, or *functionally antiferroelectric systems*). Particularly, there is much progress in engineering systems that exhibit the double-hysteresis functionality of antiferroelectrics. We will refer to those as **synthetic** (or *artificial*) **antiferroelectrics** to emphasize their distinct characteristic: in contrast to regular antiferroelectrics, which are single-phase materials, synthetic antiferroelectrics are represented by composite or multi-phase systems, such as $PbTiO_3/SrTiO_3$ superlattices, paraelectric + ferroelectric mixtures (e.g., Sm:$BiFeO_3$), and heterostructures.

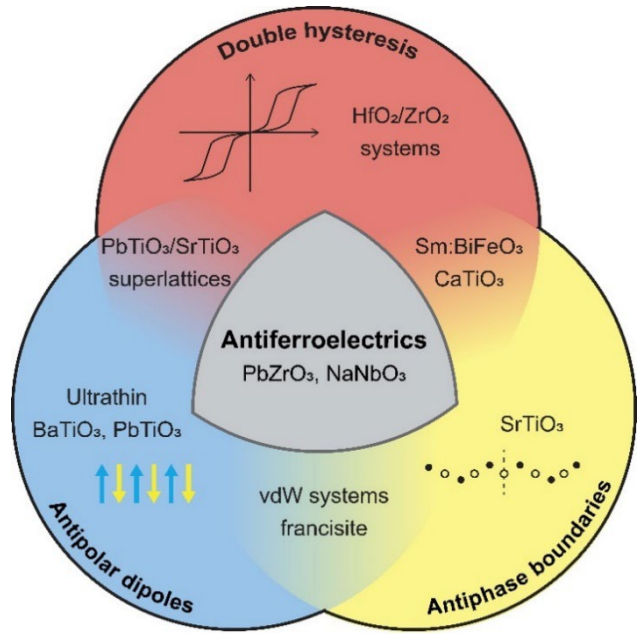

Fig. 2 | **Venn diagram illustrating the key features of antiferroelectricity and representative examples of materials that incorporate all (antiferroelectrics) or only some of them (pseudo antiferroelectrics).** Conventional antiferroelectrics, such as $PbZrO_3$ and $NaNbO_3$, exhibit antipolar dipole arrangements, antiphase boundaries and double hysteresis loops. In contrast, pseudo antiferroelectrics exhibit only some of these characteristics. For example, non-centrosymmetric layered van der Waals (vdW) materials (e.g., $CuInP_2Se_6$ and $In_2Se_3$) and engineered 2D moiré heterostructures, develop periodic antipolar structures with zero net polarization[28,29] but cannot exhibit double hysteresis loops, as no reversible transition between the nonpolar and polar states can be induced by voltage. For francisite, double hysteresis may yet be achieved, but has not been reported. Due to interfacial depolarization fields, thin films of typical ferroelectrics, such as $PbTiO_3$ and $BaTiO_3$, may split into antiparallel nanodomains that average the polarization to zero[30,31] but exhibit regular hysteresis and no antiphase boundaries. Periodic antiparallel domains (180° domains) have also been reported in ferroelectric-paraelectric superlattices of $PbTiO_3/SrTiO_3$[32] that, additionally, show double hysteresis, but no antiphase boundaries. Some fluorite-structured ferroelectrics of the $HfO_2/ZrO_2$ family[33,34] show double hysteresis loops due to the field-induced transition between a non-polar (but not antipolar) tetragonal phase and polar orthorhombic phase, which does not meet the defining criteria of antiferroelectricity. Antiphase boundaries separating structural domains (e.g., sequences of octahedral tilts) have been reported in various antiferrodistortive oxides[35], including $SrTiO_3$ at low temperature [36,37]. Some, like doped $BiFeO_3$ near the morphotropic boundary between the polar and non-polar phases[38], or $CaTiO_3$ at high electric fields[39], may also display double hysteresis loops, but the dipoles of the induced polar phase are not antiparallel in the ground state. Note that the materials included in this Venn diagram are only representative examples; the list is not exhaustive.

**Domains and Domain Walls**

In ferroic materials, energy-equivalent regions with different orientation of the order parameter are called *domains* and their boundaries *domain walls*. Antiferroelectrics also possess domain walls, at which the dipole sequence changes, for example, from "up-down-up-down" to "down-up-down-up". Notice that all the dipoles become inverted by the presence of the domain wall, even if the overall polarization is the same; such domain walls are called *antiphase boundaries* (Fig. 1), by analogy with half-unit cell discontinuities in other periodic systems[40]. The concept of antiphase boundaries is generalized as *translational boundaries* (Box 1) in systems with more complex modulations.

**BOX 1: Translational boundaries in antiferroelectrics**

Let us describe the antiferroelectric dipole modulation as a spatial *polarization wave* along the x-axis, $P = P_0 \sin\left(\frac{2\pi}{a} x + \phi\right)$, where $a$ is the modulation periodicity of the dipoles (the antiferroelectric unit cell), $P_0$ is the modulation amplitude (the maximum dipole size), and $\phi$ is a spatial phase. A translational boundary is defined as a discontinuity in the value of the spatial phase (a phase shift of the polarization wave), $\Delta\phi$. If $N$ is the number of dipoles contained within the antiferroelectric unit cell, the antiferroelectric periodicity can be expressed as $a = Nb$, where $b$ is the distance between consecutive dipoles, which typically coincides with unit cell size of the higher symmetry (paraelectric) phase (rare exceptions include materials like the hexagonal polymorph of $BaTiO_3$[41,42]). Above the antiferroelectric transition, the antipolar modulation (and any eventual translational boundary) must disappear, and the only periodicity that remains is the paraelectric one. The phase shifts that preserve the continuity of the paraelectric lattice are given by

$$\Delta\phi = 2\pi \frac{nb}{a}$$

where $n \leq N$. In the simple case of an "up-down" modulation ($N = 2; \frac{b}{a} = \frac{1}{2}$), the only non-trivial phase shift is $\Delta\phi = \pi$, defining the *antiphase boundary*. If $N = 4$ (as for $PbZrO_3$), there are three distinct translational boundaries. Longer modulations can have even more, with incommensurate antiferroelectrics having no restriction on $\Delta\phi$. Notice that, for $n = N$, the original phase is recovered, while for $n < N$ removing a translational boundary implies changing the entire dipole sequence of one of the domains. For this reason, such translational boundaries are said to be topologically protected.

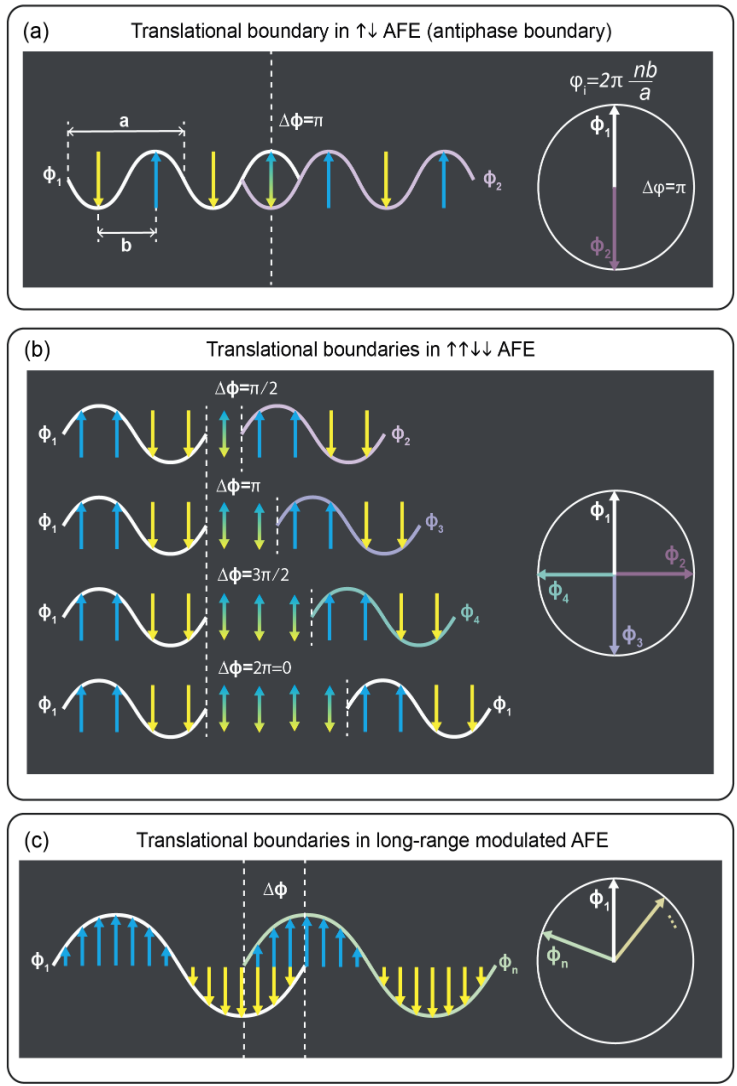

**Box figure. Antiphase boundaries and translational boundaries.** (a) Schematic of the simplest possible translational boundary: an antiphase boundary in an idealized antiferroelectric (AFE) with "up-down" (↑↓) polarization. (b) Translational boundaries in an antiferroelectric where the dipole modulation is "up-up-down-down" (↑↑↓↓;). (c) Long-range modulations allow for many phase differences between adjacent domains. Unit circles on the right illustrate mathematically possible phase differences at the translational boundaries for each type of antiferroelectric.

In the classic antiferroelectric PbZrO$_3$, the dipole sequence is "up-up-down-down", and a translational boundary may shift the dipole sequence by one, two, or three dipoles (four recovers the initial sequence), equivalent to shifting the polarization wave by $\pi/2$, $\pi$, or $3\pi/2$. Such translational boundaries are inherently polar (as they disrupt the antipolar compensation), atomically thin, and topologically protected[43–46].

In incommensurate antiferroelectrics, by definition, there is no rational relation between the periodicity of the dipoles and that of the underlying crystal lattice and, therefore, phase shifts of any value can arise at translational boundaries, analogous to domains in magnetic spin-spiral systems[47,48]. Translational boundaries in incommensurate antiferroelectrics remain unexplored.

## ANTIFERROELECTRICS OLD AND NEW

### New Phases in the Oldest Antiferroelectric

Lead zirconate, PbZrO$_3$, was the first confirmed antiferroelectric[13] and remains a paradigm for antiferroelectricity, even though it does not quite fit the ideal Kittel-like picture of "up-down" oriented dipoles with only half of them switching upon voltage application. Instead, its dipole modulation is "up-up-down-down", and upon application of an external field all the dipoles reorient from antiparallel along the pseudocubic [110] axis, to parallel pointing in the [111] direction. Concomitant with this reorientation, there is a change in crystal class, from orthorhombic (Pbam)[49] to rhombohedral (R3c).

Intermediate incommensurate phases have been reported to bridge the polar and antipolar states[50]. Moreover, even the ground state (i.e., the most stable phase calculated at 0 K) of PbZrO$_3$ is contested, with theoretical proposals ranging from antiferroelectric, with a larger (80 atoms) unit cell[51], to ferrielectric (uncompensated antipolar order), with two dipoles up and one down with orthorhombic symmetry (Ima2)[52]. A phase transition from the room-temperature Pbam symmetry to a different symmetry at low temperature has never been confirmed in bulk, although there is evidence for low-temperature polarization in field-cooled PbZrO$_3$[53,54]. Interestingly, the theoretically predicted ferrielectric phase has the same crystal structure as the translational boundaries in PbZrO$_3$[45], suggesting that such boundaries can act as nucleation points for a hypothetical/frustrated transition to a low-temperature ferrielectric phase[55]. For thin films of PbZrO$_3$, the phase diagram is further complicated by substrate clamping and/or surface tension, which induce ferroelectricity[56–60]. Substrate clamping also affects the switching speed, which is faster in free-standing membranes than in epitaxial films[61].

### Non-Collinear Systems

Recent research has revealed that ferroelectrics can develop complex polar structures beyond Ising-type collinear states, including vortex-like textures, incommensurate order, Bloch- and Néel-type walls[62,63], as well as skyrmions and hopfions[27,64,65], which are surprisingly similar to the respective non-collinear spin orderings in magnetic materials. In the 1970's, Zheludev pointed out that antiferroelectricity can exist in polar materials, provided the antipolar order occurs in the direction perpendicular to the polar axis[66] (see the *canted* state in Fig. 1). This symmetry-based observation is important, as it challenges the statement in Kittel's seminal work that the "antiferroelectric state will not be piezoelectric", allowing a material to be simultaneously antiferroelectric and polar[25].

One material that falls into this category is the improper ferroelectric $Gd_2(MoO_4)_3$. As pointed out by Jeitschko, most of the electric dipole moments in the unit cell cancel out, but due to their canting, a small spontaneous polarization of 0.2 µC/cm$^2$ arises[67]. Despite more than 50 years of research on $Gd_2(MoO_4)_3$, however, little is known about the coupling between the polar and antipolar orders, the impact of the antipolar dipole texture at the level of the domains and domain walls, or the potential existence of double hysteresis.

In magnetic systems, non-collinear orderings often arise from the so-called Dzyaloshinskii-Moriya interaction (DMI), proportional to the cross-product between adjacent spins[68,69]. In analogy to the magnetic DMI, an electric DMI has been proposed for electric dipoles that is allowed by symmetry and promotes a canting and, hence, noncollinear arrangements[70,71]. This idea provides a new pathway for non-collinear antiferroelectricity. However, symmetry considerations alone only guarantee that DMI-like interactions between electric dipoles can exist, but their magnitude (or even the actual physical mechanism that would enable them) remains unknown. Nevertheless, non-collinear polar orders have been observed and attributed to DMI-like interactions[72,73].

Frustration can also drive a system away from simple anti-ferroic order, as reflected by magnets with competing antiferromagnetic nearest and next-nearest exchange interactions, which naturally develop long-range modulated spin structures[74]. Analogously to magnetic interactions, phonon modes (e.g., polar and antipolar) may compete and lead to periodic commensurate or incommensurate order of electric dipole moments. The latter potentially plays a role in the uniaxial hyperferroelectric $Pb_5Ge_3O_{11}$, where incommensurate antiferroelectric order was observed at the ferroelectric domain walls[75-77]. A more usual type of frustration in ferroelectrics is caused by imperfect screening of depolarization fields, which typically results in the formation of antiparallel domains or, in nanoscopically confined geometries, non-collinear dipolar textures where the net polarization is averaged to zero[30,32,65].

These different examples of *beyond-Kittel* antiferroelectrics highlight the importance of local symmetry-breaking and frustration at the unit cell level. When low local symmetry (e.g., at domain walls or interfaces in heterostructures) is combined with unusual driving forces for polar order, such as improper or hyperferroelectricity, unusual antiferroelectric phenomena may arise, opening a versatile playground for novel types of non-collinear antipolar order.

**Fluorite-Structured Antiferroelectrics**

The emergence of fluorite-structured ferroelectrics, represented by the families of $HfO_2$ and $ZrO_2$ binary oxides, is marked by the discovery of a polar phase in $SiO_2$-doped hafnia thin films[78]. In addition to a range of characteristics atypical for ferroelectrics, such as polarization enhancement at reduced dimensions, some hafnia-based compounds exhibit double hysteresis[10]. A gradual transition from ferroelectric to antiferroelectric behavior upon increasing Si concentration was already reported in the first paper on ferroelectricity in hafnia[78]. The underlying microscopic mechanism, however, is not yet fully understood. A commonly accepted model attributes the double hysteresis to the field-induced transition between a non-polar tetragonal phase and a ferroelectric orthorhombic phase[15,17]. Note that the non-polar phase is not antipolar in the sense of our proposed definition (i.e., it does not present local electric dipoles that switch into the polar state). Hence, based on the above definition, $Si:HfO_2$ films with double hysteresis should be categorized as pseudo antiferroelectric.

Having said this, some hafnia-zirconia compounds can be stabilized in orthorhombic states with actual antiparallel modulations of the usual ferroelectric phase[79–82], displaying what can be interpreted as antiphase boundaries. First-principles calculations predict that such antipolar states are competitive with the polar polymorph and even lower in energy in some cases, thus fulfilling a necessary condition for the antiferroelectric behavior[83]. Further, all such polymorphs (including the monoclinic ground state) can be linked to a common parent structure via (anti)polar distortions belonging to related phonon bands[84], which would constitute a textbook example of antiferroelectric behavior involving spin-like localized electric dipoles[79,84]. Yet, as far as we know, durable double-hysteresis loops have not yet been observed in these phases, placing them, at least provisionally, in the pseudo antiferroelectric category, albeit of a different kind than the tetragonal polymorph mentioned above.

**Layered Structures**

Various physical mechanisms can lead to antiferroelectric-like responses in the multiphase systems comprised of non-antiferroelectric materials (Fig. 3). As mentioned above, these systems are termed as *synthetic antiferroelectrics* in analogy with synthetic antiferromagnets[85–87].

In ferroelectric thin films and heterostructures, depolarizing fields can induce polarization textures resembling an antiferroelectric state[75,83], with dipole configurations ranging from periodic 180° domains and flux-closing domain patterns to polar vortices, merons, skyrmions and hopfions[27,65,88–92] (Fig. 1). These structures do not exhibit net polarization at remanence, and their transition to a polarized state under external voltage can display a double hysteresis loop[32,93,94], thus fulfilling two requisites of antiferroelectrics.

Strain can also be utilized to achieve antiferroelectric-like functionality. In $PbTiO_3/SrTiO_3$ multilayers, tensile epitaxial strain favors in-plane polarization[95], and application of a vertical electric field can then switch transiently the polarization from in-plane to out-of-plane, thus generating a double-hysteresis loop for the out-of-plane polarization[96]. Choosing different strains and electrode configurations, the axes of the spontaneous and induced polarization can be swapped to generate a double hysteresis loop in the in-plane direction instead[97,98].

A third degree of freedom is interfacial chemistry or built-in fields, which can control the polarization direction in epitaxial heterostructures[99,100] to achieve opposite polarization directions perpendicular to the interface[101–104]. Finally, antipolar electric dipole ordering can emerge spontaneously in layered materials, such as those considered for ferroelectric memories in the late 50s[105]. The layered ferroelectric polarization of Aurivillius phases, for example, exhibits such antipolar ordering in the direction orthogonal to the structural planes[72,106,107]. Aurivillius phases can be epitaxially combined with perovskites[72,108], thus providing alternative pathways for mimicking antiferroelectrics.

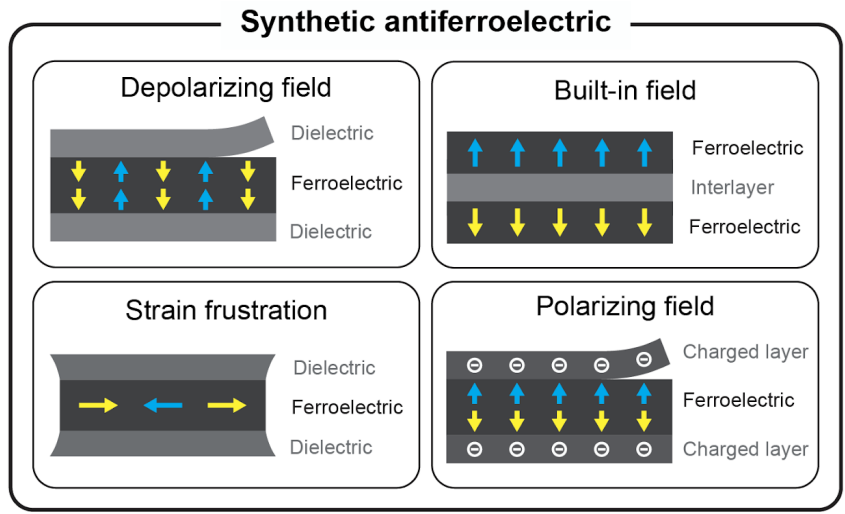

Fig. 3 | **Synthetic antiferroelectric materials using thin film architectures.** *Depolarizing field:* The depolarization effect in ferroelectric/dielectric superlattices stabilizes an antipolar dipole configuration in the ferroelectric layers at remanence, which may transition into a non-remanent polar ordering upon electric field application. *Strain Frustration:* The non-remanence of out-of-plane ferroelectric ordering can also be achieved with tensile epitaxial strain, which favors in-plane polarization. *Built-in field:* Atomic surface termination and/or interfacial band bending generate interfacial biases that favor local dipole alignment. By alternating interfacial atomic terminations in thin film heterostructures, it is possible to create multi layers exhibiting an antipolar integral configuration at remanence. Note that the interlayer involved in the termination control does not need to be dielectric, although for the purpose of achieving hysteresis loops without leakage it is desirable. *Polarizing field:* Adsorbed charge, surface off-stoichiometry, charged sheets in layered compounds, or interfacial band bending may locally bias opposite interfaces of ferroelectric thin films in opposite directions and, hence, induce an antipolar ordering at remanence.

The examples above come mostly from perovskite thin films. However, there is growing interest in layered van der Waals (vdW) materials, where weak interlayer bonding enables different stacking orders. Several vdW materials have been theoretically predicted or experimentally demonstrated to be ferroelectric[109]. In some cases, such as in $CuInP_2Se_6$, depolarizing fields may stabilize antiparallel dipole ordering and thus an antiferroelectric-like state with zero net polarization[29]. Moreover, in engineered moiré heterostructures, mechanical sliding of adjacent layers changes the arrangement of atomic columns along the out-of-plane direction, creating alternating antipolar and polar orders[28,110–113]. In the absence of external force, however, the bilayers remain in their sliding-induced states rather than returning to the initial one, thus failing a basic tenet of antiferroelectricity.

## OUTLOOK

### Thin Films

In ferroelectric thin films, depolarizing fields tend to suppress ferroelectricity, often favoring antiparallel polarization as discussed previously. Although antiferroelectrics are macroscopically not polar and thus theoretically insensitive to depolarization fields, they can also display size effects – in the opposite direction: some antiferroelectrics become ferroelectric at low thickness[114], with explanations typically involving epitaxial strain and/or built-in fields[115,116]. Recent theoretical works[60] and experiments on substrate-free membranes[117], however, show that epitaxial strain is not required to induce ferroelectricity. In fact, epitaxial strain engineering may even be used to boost the functional properties of antiferroelectrics, such as their electromechanical response[118].

Antiferroelectric membranes and strain engineering of epitaxial antiferroelectrics are likely to remain active areas of research, aiming to clarify the relative importance of strain/surface tension and depolarizing fields in the antiferroelectric size effect, and to exploit new mechanical or stacking degrees of freedom to achieve new or improved functionalities.

**Ferrielectricity**

As discussed throughout this Perspective, there are many antipolar arrangements besides the simplest "up-down-up-down" order. Some of these are uncompensated, either because there is a non-collinear order with a polar axis perpendicular to the antipolar one (see the discussion of *hybrid orders* below), or because the antiparallel dipoles differ in magnitude or number, representing a dipolar equivalent of ferrimagnets, i.e., *ferrielectrics*. Ferrielectricity was proposed by Pulvari 65 years ago[119], and subsequent research focused mostly on liquid crystals[120,121] and organic crystals[122], but recently it has been sighted in $PbZrO_3$[45,52,59].

Since ferrielectric unit cells are polar, one might ask what distinguishes them from a ferroelectric phase or from a biased antiferroelectric state. The answer is switching. Ferroelectrics have a single hysteresis loop between the +P and -P states. Antiferroelectrics have a double hysteresis loop between the non-polar ground state and the ±P field-induced polar phases. And ferrielectrics should have a triple hysteresis loop between the two polarities of the uncompensated ground state, and the two polarities of the field-induced saturated polar state. Or, to borrow the terminology of Fukuda *et al.*[123], ferroelectrics are bistable, antiferroelectrics are tristable, and ferrielectrics are tetrastable (or have a free-energy quadruple-well[124]). Such loops are rarely reported, yet they are central to the definition of ferrielectricity. Interestingly, triple hysteresis loops have been reported for $PbZrO_3$[53], although the authors did not relate them to ferrielectricity.

**Non-Collinear Hybrid Orders and Topological Singularities**

A coexistence of antiferroelectricity and ferroelectricity along different axes used to be mainly a theoretical concept, but a canted antiferroelectric order has recently been observed in potassium niobate borate ($K_3[Nb_3O_6(BO_3)_2]$), which displays both antiferroelectric characteristics (double-hysteresis loop[125] and antiphase domain walls) and ferroelectric responses (piezoelectricity and switchable domains[126]). Together with $Gd_2(MoO_4)_3$, these are just two incipient examples of the broader class of hybrid ferroelectric-antiferroelectric materials. From a symmetry perspective, it can be argued that there are several polar point groups (e.g., mm2 and 2)[66] that allow for such hybrid order, giving guidelines where to search for novel types of antipolar materials that combine phenomena traditionally viewed as mutually exclusive.

Furthermore, in close analogy to magnetism[127], long-range modulated non-collinear textures can arise (Fig. 1). Recent work on $SrTiO_3/PbTiO_3/SrTiO_3$ heterostructures[128] further emphasized the role of incommensurabilities in connection with such long-range modulated electric dipole order. Incommensurabilities, for example, lift symmetry restrictions for the phase jumps occurring at translational boundaries (Box 1), substantially increasing the number of possible domain wall states. Importantly, it has become clear that complex topologies can evolve into long-range modulated systems, and modulations of the polar and antipolar orders can coexist, exhibiting different periodicities along different axes. This possibility can be leveraged to artificially design new hybrid ferroelectric-antiferroelectric sytesm. Conversely, while ferroelectrics can have polarization vortices that can be harnessed to create antipolar textures, the question of whether vortices and related topological textures can exist in antiferroelectrics is only starting to be addressed[129]. All of which is to say that the intersection between topology and antiferroelectricity is an emergent area with a long road ahead.

**Materials Design and Practical Challenges**

The search for materials with new or improved properties is guaranteed to continue. The search is already on along several fronts outlined in this Perspective: different types of antiferroelectric order (ferrielectric, non-collinear, improper), different properties (e.g., antiferroelectric multiferroics, antipolar metals), improved functionalities (smaller hysteresis and critical fields), or improved chemistry (lead-free compositions for energy storage capacitors, complementary metal-oxide semiconductor (CMOS) compatibility for electronic devices). The lead-free search is mainly focused on perovskite niobates[130,131] (whose antipolar symmetry, fun fact, was determined in the same year as the antiferroelectricity of $PbZrO_3$)[132], as well as $BiFeO_3$-based compounds and heterostructures[93,133].

Meanwhile, many of the device challenges of antiferroelectrics are similar to those faced by their ferroelectric counterparts: scalability, low leakage, high polarization in the field-aligned state, and high resistance to fatigue. Others are more specific or application dependent. For energy storage efficiency, for example, the width of the hysteresis loops needs to be reduced[134]. Proposed strategies to address these issues include doping and strain engineering[133,135,136]. Another challenge for antiferroelectrics is the integration into CMOS technology, which limits the choice of suitable materials. Due to their Si-compatibility and high dielectric constant, hafnia-zirconia-based antiferroelectrics are the prime candidates for integrated devices, such as non-volatile memory and energy storage[137].

On the fundamental front, meanwhile, the search for a *textbook* antiferroelectric continues and has led to francisite[138], although double-hysteresis has yet to be demonstrated in this material. In the meantime, $PbZrO_3$ remains the go-to material for deepening our understanding of antiferroelectricity, ranging from fundamental aspects (e.g., domain walls and topological singularities, theoretical models) to functional properties (switching dynamics, storage efficiency and so on). In spite of its central role, the full phase diagram as a function of strain, temperature, electric field, pressure and thickness remains to be unraveled – and the advent of free-standing membranes has added new mechanical degrees of freedom.

**Theory and Simulation**

Antiferroelectrics are relatively easy to treat at a phenomenological level, e.g., by using a Landau model *à la* Kittel, with competing polar and antipolar order parameters and a triple potential well in the free energy[12,22,139]. By contrast, constructing atomistic models of antiferroelectrics, and explaining the mechanisms favoring the antipolar order over polar alternatives, is proving to be a daunting task. Perovskite oxides are a paradigmatic example of such difficulties. After decades of speculation, particularly on $PbZrO_3$ (for which arguments had been put forward for[140] and against[141] a role of flexoelectricity in stabilizing antipolar order, viewed as a form of spontaneous polarization gradient), only recently have we had convincing evidence – from first-principles simulations[142,143] and spectroscopy[144] – of the key role that oxygen-octahedral rotations play in the stabilization of the antipolar cation displacements. While such insights have made it possible to construct physically meaningful effective models for materials like PbZrO$_3$[145–147] and BiFeO$_3$-based solid solutions[133,148], they have also evidenced that antiferroelectric phenomena elude a simple atomistic description.

As of today, it seems all but impossible to identify universal mechanisms and design rules at a fundamental atomistic level, even if we restrict ourselves to the family of perovskite oxides. Interestingly, our ability to simulate antiferroelectrics and predict their non-trivial behaviors (as, e.g., in nanostructured materials, as a function of temperature and external fields) will benefit from modern machine-learning techniques that allow generation of accurate and computationally-light models from an affordable number of (heavy) first-principles calculations[149]. These models can capture implicitly the subtle effects – chemical, steric, geometric, electric, elastic – favoring antipolar order. Hence, for good or bad, computational protocols might soon be able to predict (and optimize) antiferroelectric behavior even if our physical understanding of its atomistic underpinnings will continue to be limited. Indeed, high-throughput computational studies identifying new compounds are starting to appear[150], but general design rules remain elusive. Interestingly, a much simpler problem is that of optimizing synthetic antiferroelectrics – e.g., by engineering electric and elastic constraints with clearcut physical effects – where predictive atomistic simulations are expected to be very helpful[94].

**Conclusion**

If there is a message in this Perspective, it is that, even at a venerable age (74 years since Kittel's theory and Shirane's discovery), antiferroelectrics remain in a relative infancy. Such basic aspects as the definition of antiferroelectricity or the ground state of PbZrO$_3$ are still debated, nevermind size effects, topological singularities, or new materials with emergent properties. Inspiration is still derived from the more mature research areas of ferroelectricity and antiferromagnetism, and this may yet lead to new discoveries (e.g., multi-antiferroics and [151] antiferroelectric altermagnets[152], although we have also emphasized that spins and dipoles are different, and the analogy with magnetic systems should be used prudently.

As anything in its infancy, it is easy to see that antiferroelectrics are growing, but harder to forecast how they will shape up. By pointing at some current research trends and gaps in our knowledge, we hope this article can provide some perspective for this growth.

**Acknowledgements**

All authors thank Bixin Yan for figures editing and Jiali He for proof-reading the manuscript and references. A.G. and D.M. thank NTNU for support via the Lars Onsager Professorship 2024 that was awarded to A.G.; D.M. acknowledges funding from the European Research Council (ERC) under the European Union's Horizon 2020 Research and Innovation Program (Grant Agreement No. 863691) and thanks NTNU for support through the Onsager Fellowship Program and the Outstanding Academic Fellow Program. A.G. acknowledges support from the National Science Foundation (grant DMR-2419172) and Intel Corporation. M.T. acknowledges support from the Swiss National Science Foundation under Project No. 200021-231428 and the ETH Zurich Research Grant funding under reference 22-2 ETH-016. G.C. acknowledges financial support from the FET-Open Grant N° 964931 (Project TSAR, DOI: 10.3030/964931) from the European Union's Horizon 2020 research and innovation program, as well as grants 2021 SGR 0129 and PID2023-148673NB-I00 from the Catalan and Spanish national research agencies, respectively. All research at ICN2 is supported by a Severo Ochoa Grant CEX2021-001214-S. Research at LIST is supported by the Luxembourg National Research Fund (FNR) through grant C21/MS/15799044/FERRODYNAMICS.

**Author contributions**

All authors contributed equally; listed in alphabetical order.

**Competing Interests**

The authors declare no competing interests.